\documentclass[12pt]{article}
\usepackage{setspace}
\usepackage{amsmath, amssymb, amsthm,float,graphicx}
\def\be{\begin{equation}}
\def\ee{\end{equation}}

\def\be{\begin{equation}}
\def\ee{\end{equation}}

\def\bg{\bar{g}}
\def\beq{\begin{eqnarray}}\def\eeq{\end{eqnarray}}
\def\ba#1\ea{\begin{align}#1\end{align}}
\def\bg#1\eg{\begin{gather}#1\end{gather}}
\def\bm#1\em{\begin{multline}#1\end{multline}}
\def\bmd#1\emd{\begin{multlined}#1\end{multlined}}

\def\({\left(}
\def\){\right)}
\def\[{\left[}
\def\]{\right]}

\begin{document}

\title{Holographic Lifshitz fluids in 1+1 dimensions}

\author{
%{\bf {\normalsize Rabin Banerjee}$^{a,}
%$\thanks{rabin@bose.res.in}},\,
{\bf {\normalsize Arpan Bhattacharyya}$^{a}
$\thanks{bhattacharyya.arpan@yahoo.com}}~
%{\bf {\normalsize Arindam Ghosh Hazra}
%$^{a,}$\thanks{arindamg@bose.res.in}},\, 
{\bf {\normalsize and~ Dibakar Roychowdhury}
$^{b}$\thanks{dibakarphys@gmail.com}}\\
$^{a}$ {\normalsize Center for Gravitational Physics, Yukawa Institute for Theoretical Physics,}\\{\normalsize Kyoto University, Kyoto 606-8502, Japan.}\\
$^{b}$ {\normalsize Department of Physics, Swansea University,}\\ {\normalsize Singleton Park, Swansea SA2 8PP, United Kingdom}
\\[0.3cm]
%$^{c}${\normalsize }\\
%{\normalsize }\\[0.3cm]
%$^{d}${\normalsize }\\
%{\normalsize }\\[0.3cm]
}
\date{}

\maketitle
\begin{abstract}
{In this paper, using the techniques of Gauge/gravity duality we explore the hydrodynamic  regime of $ z=3 $ Lifshitz fixed points in $ 1+1 $ dimensions. The speed of sound in the non relativistic plasma turns out to be $\sqrt{3}$, which clearly violates the conjectured upper bound. We identify this as a natural consequence of the Lorentz symmetry breakdown at $ z=3 $ Lifshitz fixed point. In our analysis, we  compute  bulk viscosity to entropy ratio for  quantum field theory dual to Lifshitz gravity in $ 2+1$ dimensions and it turns out that for this particular holographic model this ratio is above the conjectured lower bound.}
\end{abstract}
\newpage

%\tableofcontents
%\tableofcontents
%%%%%%%%%%%%%%%%%%%%%%%%%%%%%%%%%%%%%%%%%%%%%%%%%%%%%%%%%%%%%%%%%%%%
%\balancecolsandclearpage
\onehalfspacing
%\date{\today}
\vskip 1cm
%%%%%%%%%%%%%%%%%%%%%%%%%%%%%%%%%%%%%%%%%%%%%%%%%%%%%%%%%%%%%%%%%%
\section {Overview and Motivation}
Fluid dynamics is the low energy effective description of finite temperature QFTs
 perturbed from their usual thermal equilibrium by means of small fluctuations \cite{LL}. In recent times, the Gauge/gravity duality has been found to shed enough light on various key aspects of hydrodynamic transports in a strongly coupled fluid \cite{Son}. In other words, the Gauge/gravity duality provides a unique framework to explore the low frequency behavior of two point correlators between different components of the stress tensor in the strong coupling regime.
 
Among various remarkable outcomes of the Gauge/gravity duality, it is the existence of the sound modes in a strongly coupled plasma \cite{Policastro} that turns out to be the prime focus of our current analysis. It is the standard lesson of fluid dynamics which tells us that sound waves can always propagate through a finite temperature fluid medium. The information regarding the sound velocity could be extracted by knowing the poles in the expression for the retarded two point correlators between some of the components of the stress tensor  \cite{Policastro}.

 The purpose of the present analysis would be to employ holographic techniques in order to compute transports, namely the bulk viscosity to entropy ratio for $(1+1)$ dimensional  QFTs dual to Lifshitz gravity in (2+1) dimensions. In our analysis we consider the bulk construction to be the $ z=3 $ Lifshitz black hole solutions of New Massive Gravity (NMG) in ($ 2+1 $) dimensions \cite{AB}. There are in fact several motivations behind our present analysis which we quote systematically.

(i) Since the Lorentz symmetry is explicitly broken for the dual fluid dynamical system under consideration,  therefore one should expect some natural violation of the conjectured upper bound on the speed of sound \cite{Cherman,Hohler} particularly in 1+1 dimensions \cite{David}. In fact, we argue that the conjectured upper bound on the speed of sound in $1+1$ dimensions is the consequence of the conformal invariance which might get violated in case of  broken scale invariance i.e, when $z\neq 1$. The present calculation is a concrete example in favour of the above argument.

(ii) It turns out to be an interesting issue to explore the effect of genuine higher derivative (NMG) corrections on the bulk viscosity to entropy ($ \zeta/s $) ratio in $ 1+1 $ dimensions and to see whether we get a bound violation just like the case for the shear viscosity to entropy ratio \cite{Brigante}\footnote{Dual CFT may or may not be a unitary one. It is most likely a non-relativistic verison of logarithimic CFT as shown in \cite{Bergshoeff1}. This kind of CFT's is also interesting though sometimes there are problems regarding unitarity as the central charge takes negative value. But still these CFT's have many applications as  interesting mathematical models as a well as describing some condensed matter systems away from unitarity  \cite{Flohr}.  In this paper we are actually making a prediction about the hydrodynamic transport of these non-relativistic logarithmic CFT's using the AdS/CFT dictionary for the first time. }. Interestingly this issue is completely unexplored for non relativistic strongly coupled fluids (particularly in lower dimensions) and the present calculation is the first non trivial example to the best of our knowledge.

(iii) Finally, and most importantly, there are enough motivations to study non relativistic fluids in $ 1+1 $ dimensions solely from the perspective of the usual condensed matter systems \cite{conmat,conmat1}.
%%%%%%%%%%%%%%%%%%%%%%%%%%%%%%%%%%%%%%%%%%%%%%%%%%%%%%%%%%%%%
\section{The bulk solution}
We start our analysis with a brief introduction to the bulk construction, namely the Lifshitz solution in the context of New Massive Gravity (NMG)\cite{Bergshoeff}. The corresponding action in the NMG reads as \cite{Bergshoeff},
\be
S_{NMG}=\frac{1}{2\kappa}\int d^{3} x \sqrt{g}\Big[R-2\Lambda-\lambda\Big(R_{\mu\nu}R^{\mu\nu}-\frac{3}{8} R^2\Big)\Big].\label{E1}
\ee
Here $\Lambda$ is the cosmological constant and $\lambda$ is the NMG coupling constant. The above action (\ref{E1}) admits a Lifshitz solution of the type \cite{AB},
\be
ds^2=- \frac{r^{2z}}{l^{2z}} f(r) dt^2+\frac{dr^2}{g(r)}+\frac{r^2}{l^2} dx^2\label{E2}
\ee
where \cite{AB},  
\begin{equation}
f(r)=\frac{r^2}{l^2} g(r)=\Big(1-\frac{ M^2 l^2}{r^2}\Big)
\end{equation}
together with, 
\begin{equation}
z=3,\quad \Lambda=-\frac{13}{2 l^2},\quad  \lambda=-2 l^2\,.
\end{equation}
The horizon is located at $r=r_{h}= M l\,.$ For convenience, we set $l=1$ in the subsequent calculations.
 In our analysis the equations of motion corresponding to the hydrodynamic (spin $ 0 $) fluctuations turn out to be fourth order linear differential equations which thereby invite a completely different as well as challenging platform to compute the hydrodynamic transports for the non relativistic $ z=3 $ plasma at strong coupling.
%%%%%%%%%%%%%%%%%%%%%%%%%%%%%%%%%%%%%%%%%%%%%%%%%%%%%%%%%%%%
\section{Thermodynamics} \label{sec3}
In this section we provide a brief discussion to the basic thermodynamic principles and/or identities which any fluid dynamical system must obey. These are the so called first law of fluid mechanics and the Euler relation \cite{LL}. Before we actually come to these issues, let us first note that the Wald entropy density ($ s $) for the above black hole solution (\ref{E2}) turns out to be \cite{AsS}, 
\be
s=\frac{8 \pi  r_{h}}{\kappa}\label{E3}
\ee
which also matches with \cite{AB1} in the appropriate limit.

The energy density ($ \epsilon $) could be formally obtained by integrating over the Noether charge associated with the time like Killing symmetry
%\footnote{One can derive this result by using several other techniques for example, using the notion of %renormalized holographic stress tensor \cite{stress, stress1} or dimensionally reducing to dilaton gravity %\cite{dilaton}. Interestingly, all these methods yield the same result and therefore are claimed to be %consistent with each other \cite{consistent}.} \cite{AB1},
\be
\epsilon = \frac{r_{h}^4}{\kappa}.\label{E4}
\ee
The Hawking temperature ($ T_h $) associated with the black hole solution (\ref{E2}) turns out to be \cite{AB}, \cite{AB1},
\be
T_{h}=\frac{r_{h}^{z}}{2\pi}.\label{E5}
\ee
It is now in fact a quite trivial task to check that the above entities (\ref{E3}), (\ref{E4}) and (\ref{E5}) trivially satisfy the first law \cite{LL},
\be
d\epsilon= T_{h} d s.
\ee
Next, we compute the pressure density ($ p $) using the Euler relation \cite{LL},
\be
p=T_{h} s-\epsilon =\frac{3 r_{h}^4}{\kappa}.\label{E7}
\ee
Finally, using (\ref{E4}) and (\ref{E7}) the velocity of sound ($ v_{s} $) turns out to be %\footnote{The velocity is measured in the unit of the speed of the light $v$ in this medium. One can measure the speed of light of a system independently and after that one will be able to fix the speed of sound in a specific unit.  Also to note that in  non relativistic systems there is no upper bound on the speed of light. A massless gauge field in the boundary of this space will propagate with this velocity ($v$).  },
\be
v_{s}=\sqrt{\frac{\partial p}{\partial \epsilon}}=\sqrt{3}.\label{E8}
\ee
Eq.(\ref{E8}) is precisely the speed of sound propagation from the perspective of the boundary hydrodynamics. The fact that $ v_s >1 $ should not come as a surprise since we are dealing with a non relativistic system corresponding to $ z=3 $ fixed point where in principle there should not be any upper bound on the speed of light. 
%Incidentally because of this reason, the (conjectured) universality bound on the charge diffusion does not %in general hold for Lifshitz like fixed points \cite{Rtiz}.

Finally, using (\ref{E4}) and (\ref{E5}) the heat capacity turns out to be,
\begin{eqnarray}
C= \frac{4}{3 \kappa }\left( 2 \pi T_{h} \right)^{1/3} \label{E9}
\end{eqnarray}
which turns out to be positive definite. Eq.(\ref{E9}) is an important ingredient in order to estimate the speed of sound ($ v_s $) inside a fluid medium \cite{kittel}.
% We will make more specific comments about it in the concluding section.

%%%%%%%%%%%%%%%%%%%%%%%%%%%%%%%%%%%%%%%%%%%%%%%%%%%%%%%%%%%%%%%%%%%%%%%%%
\section{Sound velocity in Lifshitz hydrodynamics}
The purpose of the present section is to provide a systematic derivation of the hydrodynamic dispersion relation in the context of Lifshitz hydrodynamics and to compute the speed of sound as well as the coefficient of bulk viscosity directly from the dispersion relation itself. 
 
In the standard framework of fluid dynamics, the velocity for sound modes are obtained by knowing the pole(s) in the expression for the retarded correlator(s) of the type $<T_{tx}T_{tx}>$ \cite{Kovtun, Policastro}. As far as the current analysis is concerned, the stress tensor corresponding to the boundary non conformal fluid could be formally expressed as following \cite{Hoyos} , 
\be \label{trans}
T^{\mu\nu}=(\epsilon+p) u^{\mu}u^{\nu}+p\eta^{\mu\nu}-\zeta(u^{\mu}u^{\nu}+\eta^{\mu\nu})\partial_{\lambda}u^{\lambda}+\Pi_{A}^{[\mu\nu]}+\Big(u^{\mu}\Pi_{A}^{[\nu\sigma]}+u^{\nu}\Pi_{A}^{[\mu\sigma]}\Big)u_{\sigma}.
\ee
where $u_{\mu}$ (for the present case) is the 2-velocity of the fluid satisfying $u_{\mu}u^{\mu}=-1$ and $\zeta$ is the coefficient of bulk viscosity which appears as a consequence of the broken conformal invariance in the boundary theory.  At this stage it is noteworthy to mention that for uncharged Lifshitz fluids in general there exist two transport coefficients (at the level of first order viscous hydrodynamics) namely the coefficient of bulk viscosity ($ \zeta $) and an additional transport $\alpha$ that comes from the antisymmetric part of (\ref{trans}) which could be formally expressed as \cite{Hoyos},
\be
\Pi_{A}^{[\mu\nu]}=-\alpha u^{[\mu}u^{\alpha}\partial_{\alpha}u^{\nu]}.
\ee
This additional transport appears due to the broken Lorentz boost invariance at a Lifshitz fixed point.
  
In order to find pole(s) in the expression for the retarded correlator(s), we first perturb the fluid in its rest frame as \cite{David},
\begin{eqnarray} \label{fluc}
T^{tt}&=&T^{(0)tt}+\kappa T^{(1)tt}+\kappa^2T^{(2)tt}+\cdots\nonumber\\
 T^{tx}&=&T^{(0)tx}+\kappa T^{(1)tx}+\kappa^2T^{(2) tx}+\cdots\nonumber\\
T^{xt}&=&T^{(0)xt}+\kappa T^{(1)xt}+\kappa^2 T^{(2)xt}+\cdots \nonumber\\
  T^{xx}&=&T^{(0)xx}+ \kappa T^{(1)xx}+\kappa^2 T^{(2) xx}+\cdots \nonumber\\ 
   u^{t}&=&u^{(0)t}+\kappa u^{(1)t}+\kappa^2 u^{(2)t}+\cdots\nonumber\\ u^{x}&=&u^{(0)x}+\kappa u^{(1)x}+\kappa^2u^{(2)x}+\cdots\nonumber\\
\epsilon&=&\epsilon^{(0)}+\kappa\epsilon^{(1)}+\kappa^2\epsilon^{(2)}+\cdots\nonumber\\
p&=&=p^{(0)}+\kappa p^{(1)}+\kappa^2 p^{(2)}+\cdots.\label{E10}
\end{eqnarray}

We have expanded all the quantities order by order in fluctuations around their respective equilibrium values. $\kappa$ denotes the perturbation parameter. In our analysis we keep only linear order terms in the fluctuations i.e only terms $\mathcal{O}(\kappa).$ Also $\epsilon^{(0)}$ and $p^{(0)}$ denote the zeroth order energy density and pressure and independent of $x$ and $t$.\\
Due to the presence of the constraint, $u_{\mu}u^{\mu}=-1$, one could  show that,
\be
u_{t}u^{t}+u_{x}u^{x}=u^{(0)x}u^{(0)x}-u^{(0)t}u^{(0)t}-2\kappa\, u^{(0)t} u^{(1)t}+\kappa^2(-2u^{(0)t}u^{(2)t}+(u^{(1) x})^2)=-1.
\ee
 Since we are expanding around a static equilibrium solution, therefore we can fix, $u^{(0)t}=1$ and $u^{(0)x}=0$ and it follows,
\be
u^{(1)t}=0\,, \,u^{(2)t}=\frac{1}{2} (u^{(1) x})^2.
\ee
So upto $\mathcal{O}(\kappa)$ we have $u^{t}=1+\mathcal{O}(\kappa^2).$ \\
Next we concentrate on the anti-symmetric part  of (\ref{trans}).
Using (\ref{fluc}) we get,
\be
\Pi^{[tx]}_{A}=-\frac{\alpha}{2}\Big(\kappa\partial_{t} u^{(1) x}+\mathcal{O}(\kappa^2)\Big)
\ee
where we have neglected $\delta (u^{x}\partial_{x}u^x)$  and $\delta (u^{x}\partial_{x}u^t),$ as these terms are second order in fluctuations ($\mathcal{O}(\kappa^2)$).

From (\ref{fluc}) we note,
\be
 T^{tx}=(\epsilon+p) u^{t} u^{x}-\zeta u^{t} u^{x}\partial_{\lambda}u^{\lambda}+\Pi^{[tx]}_{A}+(u^{t}\Pi^{[x\sigma]}_{A}+u^{x}\Pi_{A}^{[t\sigma]})u_{\sigma}.
\ee
which could be further expressed as keeping only linear order terms , 
\be \label{txt}
T^{(0)tx}=0\,,\,T^{(1) tx}=(\epsilon^{(0)}+p^{(0)}) u^{(1)x}-\alpha\, \partial_{t} u^{(1) x}.
\ee

Proceeding in the same way we get,
\be
 T^{(0)xt}=0,\,T^{(1)xt}=(\epsilon^{(0)}+p^{(0)}) u^{(1)x}
\ee
which solves $ u^{(1)x}$ as,
\be \label{ap1}
 u^{(1)x}=\frac{ T^{(1)xt}}{(\epsilon^{(0)}+P^{(0)})}.
\ee

Next we simplify(\ref{txt}) further which finally yields,
\be 
T^{(1)tx}= T^{(1)xt}-\frac{\alpha}{ (\epsilon^{(0)}+p^{(0)})}\partial_{t}  T^{(1)xt}-\partial_{t}\Big(\frac{\alpha}{(\epsilon^{(0)}+P^{(0)})}\Big)T^{(1)xt}.
\ee
 Notice that when $\alpha=0$ we get $ T^{(1)xt}=T^{(1)tx}$ and which is thereby consistent. The last term vanishes as the derivative of the $\epsilon^{(0)}$ and $p^{(0)}$ vanish. So,
\be \label{ap5}
T^{(1)tx}= T^{(1)xt}-\frac{\alpha}{ (\epsilon^{(0)}+p^{(0)})}\partial_{t}  T^{(1)xt}.
\ee

Furthermore, we can also write down the following two equations namely using the same logic,
\be \label{ap5a}
T^{tt}=\epsilon^{(0)}+T^{(1)tt},\quad T^{(0)xx}=0,\, T^{(1)xx}= p^{(1)}-\frac{\zeta}{(\epsilon^{(0)}+p^{(0)})}\partial_{x}( T^{(1)xt})
\ee
where we have used (\ref{ap1}) to eliminate $ u^{(1)x}$. Note that, here we have neglected all the derivatives of the thermodynamical quantities in (\ref{ap5}) and (\ref{ap5a}) as they are themselves first order in the fluctuations and thereby contribute at the level of second order in the fluctuations.

 Next,  we plug all these in the conservation equations which yields,
\be \label{ap2}
\partial_{t} T^{(1)tx}+\partial_{x}  T^{(1)xx}=0
\ee
and
\be
\partial_{x} T^{(1)xt}+\partial_{t}  T^{(1)tt}=0.
\label{e2}
\ee
We have used the fact that $\partial_{t}\epsilon^{(0)}=0.$ The zeroth order variation of energy is zero.\\
Then following \cite{David} from (\ref{ap2}) we obtain,
\be \label{ap6}
\partial_{t} T^{(1)tx}+\frac{\delta p^{(1)}}{\delta \epsilon^{(1)}}\partial_{x} T^{(1)tt}-\frac{\zeta}{(\epsilon^{(0)}+p^{(0)})}\partial^{2}_{x}(T^{(1)xt})=0.
\ee
We have used the fact $T^{(1)tt}=\epsilon^{(1)}$ and used the chain rule.\\
In the next step,  using (\ref{ap5}) we first replace $T^{(1)tx}$ in (\ref{ap6}) and then take the Fourier transform of the above set of equations (\ref{ap2})  and (\ref{e2}) which finally yields,
\be \label{ap4}
i\, q \partial_{\epsilon} p  T^{(1)tt}+\Big(-i \,\omega+\frac{\omega^2\alpha}{(\epsilon^{(0)}+p^{(0)}}\Big) T^{(1)xt}+q^2\, \frac{\zeta}{(\epsilon^{(0)}+p^{(0)})} T^{(1)xt}=0
\ee
and
\be \label{ap3}
i\,q  T^{(1)xt}-i\, \omega  T^{(1)tt}=0.
\ee
Using (\ref{ap3}) one can replace $T^{(1)tt}$ in (\ref{ap4}) to obtain,
\be
T^{(1)tt}=\frac{q}{\omega}  T^{(1)xt}.
\ee
So we have,
\be \label{ap11}
i \frac{q^2}{\omega} \partial_{\epsilon} p  T^{(1)xt}+\Big(-i \,\omega+\frac{\omega^2\alpha}{(\epsilon^{(0)}+p^{(0)})}\Big)T^{(1)xt}+q^2\, \frac{\zeta}{(\epsilon^{(0)}+p^{(0)})} T^{(1)xt}=0
\ee
We futher simplify to obtain,
\be \label{ap7}
\Big(\omega^2-\, q^2\partial_{\epsilon}p+\frac{i\,q^2\,\omega\,\zeta}{(\epsilon^{(0)}+p^{(0)})}+\frac{i\,\omega^3\alpha}{(\epsilon^{(0)}+p^{(0)})}\Big) T^{(1)xt}=0.
\ee
Our next task would be to solve this equation for the frequency $\omega$ perturbatively upto $\mathcal{O}(q^2)$, where $ q $ is the spatial momentum. For this we make the  following ansatz for $\omega$ in terms of $q$ to get the dispersion relation of the following form, 
\be \label{ap8}
\omega= v_{s} q-i\,\frac{\Gamma_{s}}{2} q^2.
\ee
We plug this into  (\ref{ap7}) and match the coefficients of $q^2$ term to obtain,
\be
v_{s}^2=\partial_{\epsilon}p.
\ee
which is the speed of sound in the (Lifshitz) fluid medium. Next, we match the coefficients of $q^2$ terms in order to obtain the damping constant,
\be
\Gamma_{s}=\frac{\zeta+v_{s}^{2}\alpha}{(\epsilon^{(0)}+p^{(0)})}.
\ee
 
 Finally, using the so called Euler relation,
\be
\epsilon^{(0)} + p^{(0)}=T_{h} s
\ee
one can in fact express (\ref{ap8}) in a more sophisticated form as,
\be \label{dis}
\omega= v_{s} q-i \Big(\frac{\zeta+v_{s}^2 \alpha}{2 T_{h} s}\Big)q^2
\ee
which immediately suggests that for Lifshitz like fluids one should be able to define an effective bulk viscosity $\zeta_{lif}$ as,
\be
\zeta_{lif}=\zeta+v_{s}^2\alpha
\label{e1}
\ee
such that,
\be
\Gamma_{s} T_{h}=\frac{\zeta_{lif}}{s}.
\ee
Eq.(\ref{e1}) is one of the major observations of the present paper. It immediately suggests that for Lifshitz like fluids the effective value of the bulk viscosity is enhanced due to the broken Lorentz invariance. Therefore, it also confirms the fact that the bulk viscosity to entropy ratio ($ \zeta/s $) for Lifshitz like systems should also be above the conjectured lower bound \cite{David}. Our goal for the next section would be to compute this from the gravity calculations in the bulk, where we will consider a particular model of $ z=3 $ Lifshitz black holes in ($ 2+1 $) dimensions \cite{AB}.

%With the above result in hand, our next task is twofold. First, to check the validity of the conjectured upper bound on the velocity of sound \cite{Cherman, Hohler} for non conformal fluids corresponding to $ z=3 $ Lifshitz fixed point and second, to explore the effect of \textit{non perturbative} higher derivative (NMG) corrections on the effective bulk viscosity to entropy ($ \zeta_{lif}/s $) ratio in $ 1+1 $ dimensions.
% and comment on its conjectured lower bound \cite{Buchel,David}.

%%%%%%%%%%%%%%%%%%%%%%%%%%%%%%%%%%%%%%%%%%%%%%%%%%%%%%%%%%%%%%%%%%%%%%
\section{Dispersion relation and holography}
The first step towards computing the holographic sound modes is to turn on the following (with respect to boundary rotational symmetry) perturbations in the bulk namely \cite{Policastro},
\begin{eqnarray}
h_{\mu\nu}= \Big\{h_{tt},h_{tx},h_{xx}\Big\}
\end{eqnarray}
where we have fixed the gauge $ h_{\mu r} =0$. This gauge fixing does not fully exhaust all the gauge freedoms of the full theory and one is still left  with the residual gauge freedom under the diffeomorphism $x^\mu \rightarrow  x^\mu + \xi^\mu , \mu = (t,r,x)$. As a consequence of this the equations of motion corresponding to the above fluctuations exhibit residual gauge and/or diffeomorphism invariance  ($ \delta g_{\mu \nu}=\nabla_{\mu}\xi_{\nu}+ \nabla_{\nu}\xi_{\mu}$) \cite{Policastro}.  As a result the most general gauge invariant fluctuation turns out to be,
\begin{eqnarray}
\mathcal{Z}&=-c_{1}(r) H_{tt}(r)+c_{2}(r) H_{tx}(r)+c_{3}(r) H_{xx}(r)\nonumber\\&-q^2 \Big(r^{2z}f(r) H_{tt}(r)\Big)^{'}+2 q \omega \Big(r^2 H_{tx}(r)\Big)^{'}\nonumber \\&+ \omega ^2 \Big(r^2 H_{xx}(r)\Big)^{'}\label{E16}
\end{eqnarray}
where the individual coefficients ($ c_i $) turn out to be,
%\begin{widetext}
\be \nonumber
 c_{1}(r)= \frac{\left(q^2 r^3 ( M-r) ( M+r) \left(q^2 r^2 \left(6  M^2-23  M r^2+18 r^4\right)+ \omega ^2 \left( M-2 r^2\right)\right)\right)}{\left( M-r^2\right) \left( \omega ^2+q^2 r^2 \left(2  M-3 r^2\right)\right)}\ee
\be \nonumber
 c_{2}(r)=2  q r^2 \omega  \left(\frac{r}{ M-r^2}-\frac{4 q^2 r \left( M-3 r^2\right)}{ \omega ^2+q^2 r^2 \left(2  M-3 r^2\right)}-\frac{1}{r}\right)
\ee
\be 
 c_{3}(r)=\frac{ \left( r \omega ^2 \left(q^2 r^2 \left(6  M^2-23  M r^2+18 r^4\right)+ \omega ^2 \left( M-2 r^2\right)\right)\right)}{\left(r^2- M\right) \left( \omega ^2+q^2 r^2 \left(2  M-3 r^2\right)\right)}.
\ee
%\end{widetext}

In order to arrive at (\ref{E16}) we have used the following metric redefinitions namely,
\begin{eqnarray}
h_{xx}&=& e^{-i \omega t+ i q x} r^2 H_{xx}(r)\nonumber\\
h_{tx}&= &e^{-i \omega t+ i q x} r^2 H_{tx}(r)\nonumber\\
h_{tt}&=& -e^{-i \omega t+ i q x}r^{2z} f(r) H_{tt}(r).
\label{eh}
\end{eqnarray}
Note that here prime denotes derivative with respect to the radial coordinate ($ r $). 

Using the linearized equations of motion of individual scalar perturbations, it is indeed quite interesting to note that unlike the previous examples in the literature \cite{Policastro}, the gauge invariant combination (\ref{E16}) satisfies a sixth order linear differential equation of the type,  
\begin{eqnarray} \label{master}
&A_{0}(r,\omega,q) \mathcal{Z}^{''''''}+A_{1}(r,\omega,q) \mathcal{Z}^{'''''}+A_{2}(r,\omega,q) \mathcal{Z}^{''''}+\nonumber\\&A_{3}(r,\omega,q)\mathcal{ Z}^{'''}+A_{4}(r,\omega,q)\mathcal{ Z}^{''}+A_{5}(r,\omega,q) \mathcal{Z}^{'}+\nonumber\\&A_{6}(r,\omega,q) \mathcal{Z}=0\label{E19}
\end{eqnarray}
where details of the coefficients ($ A_i $) are too cumbersome and not quite illuminating. However, in the following we explain in details how one could in principle arrive an equation of the above type (\ref{E19}). Using (\ref{E16}), we first take a linear combination of various powers of derivatives of  $ \mathcal{Z} $ (with some arbitrary coefficients $A_{i}(r)$) starting from the sixth derivative level. This will contain derivatives of various powers of $ H $'s. Now one uses the equations of motion for $ H$'s and replace all the higher order derivatives in terms of lower order derivatives. Finally, we end up an equation which contain three second order derivatives in the metric, two first order derivatives in the metric and three metric coefficients with zero derivative level. All of these (altogether) seven different metric components and their derivatives will be associated with some specific combination of  the coefficients $ A_{i} $'s. Each of those combinations has to vanish in order to satisfy  (\ref{E19}) which yield seven algebraic relations for $ A_{i} $'s that determine them uniquely.

Next, we explore (\ref{E19}) in two asymptotic regions namely, near the boundary ($ r\rightarrow \infty $) of the spacetime and at the horizon ($ r=r_h=1 $).
We first consider the following ansatz for $\mathcal{Z}$,
\be \label{z}
\mathcal{Z}=\left(1-\frac{1}{r}\right)^{\alpha}F(r).
\ee
Considering the near horizon expansion and considering the ingoing wave boundary condition \cite{Policastro}, the parameter $\alpha$  could be read off as,
\begin{eqnarray}
\alpha=-\frac{i \omega}{4\pi T_{h}}=-\frac{i \omega}{2}.
\end{eqnarray}

Next, considering the large $ r $ limit we expand $F(r)$ in the frequency ($ \omega $) as well as in the momentum ($q$),
\be
F(r)=F_{0}(r)+ \omega F_{1}(r)+ q\, F_{2}(r)+\omega^2 F_{3} (r)+ \omega\, q F_{4}(r)+ q^2 F_{5}(r)
\ee
and extract the finite piece in the limit $r\rightarrow \infty$. Keeping those finite pieces intact and considering the asymptotic normalization condition \cite{Policastro},
\be
F(\infty)=0\label{E24}
\ee
we get the following  quadratic equation for $\omega$,
\be \label{dis1}
-\frac{\omega ^2 (115 i \, e_{1}-160 i \, e_{2}+288 i \, e_{3}-240 i \, e_{4}+60\, q)}{96 \,q}-3 i\, \omega=0.
\ee
Solving (\ref{dis1}) equation upto $\mathcal{O}(q^2)$ we finally arrive at the cherished dispersion relation of the following form, \be
\omega= v_{s}q-\frac{i}{2} \Gamma_{s} q^2 \label{E25}
\ee
where,
\be \label{vs}
v_{s}=-\frac{288}{(115 e_{1}+16 (-10  e_{2}+18 e_{3}-15 e_{4}))},
\ee
and
\be \label{ap10}  \Gamma_{s}=\frac{34560}{(115 e_{1}+16 (-10 e_{2}+18 e_{3}-15 e_{4}))^2}.\ee
$e_{1}, e_{2}, e_{3}$ and $e_{4}$ are the four left over integrating constants.  Combining (\ref{vs}) and (\ref{ap10}) it is evident that \footnote{ We thank anonymous referee for pointing this out to us.} ,
\be \label{vsref}
\Gamma_{s}=\frac{5}{12}v_s^2.
\ee
So this suggest that $\Gamma_{s}$ and $v_s$ are not independent.

 As the gauge invariant combination satisfies a six order differential equation (\ref{E19}) and there are in fact two boundary conditions namely, the in going wave boundary condition and the asymptotic boundary condition (\ref{E24}), therefore in principle we can uniquely fix two of the six unknown coefficients. Therefore, finally we are left with four undetermined constants those appear above in the expressions for $ v_s $ and $ \Gamma_{s} $.
Next, we compare (\ref{E25}) with the dispersion relation previously obtained in (\ref{ap8}). In (\ref{ap8}) we have identified $v_{s}^2=\partial_{\epsilon}p$ as the speed of sound which is computed in (\ref{E8}) using the gravity solution. So equating (\ref{vs}) and (\ref{E8}) one can uniquely fix the combination of the remaining four constants as, 
\be \label{ap9}
(115 e_{1}+16 (-10  e_{2}+18 e_{3}-15 e_{4}))=-\frac{288}{\sqrt{3}}.
\ee
Surprisingly once this special combination of four integrating constants get (uniquely) fixed one can uniquely calculate the $\Gamma_{s}$.
Using (\ref{ap10}) and (\ref{ap9}) we finally obtain,
\be
\Gamma_{s}=\frac{5}{4}
\ee
which finally yields,
\be
\Gamma_{s} T_{h}=\frac{\zeta_{lif}}{s}=\frac{5}{24 \pi} v_{s}^2 \approx 0.1989 > \frac{1}{4 \pi} \label{E26}
\ee
which clearly seems to preserve the conjectured lower bound \cite{ David} as we expected from our analysis in the previous section.  %We have used the values of $T_{h}$ and $v_{s}$ as calculated in the Section~(\ref{sec3}).

%Eq.(\ref{E26}) represents the full \textit{non perturbative} NMG correction to $ \zeta/s $ ratio in $ 1+1 $ dimensions. In \cite{Brustein:2009tk}, it has been argued for the first time that the presence of effective coupling for gravitons could modify the sound damping coefficient. However, the arguments of \cite{Brustein:2009tk} are mostly valid for conformal case, whereas on the other hand, our analysis might be regarded as the first non trivial example in the context of non conformal fluids.
%%%%%%%%%%%%%%%%%%%%%%%%%%%%%%%%%%%%%%%%%%%%%%%%%%%%%%%%%%%%%%%%
\section{Summary and final remarks}
Let us now summarize the key findings of our analysis. The present analysis has two major outcomes. The first observation tells us that for this particular holographic model under consideration the speed of sound propagation ($ v_s $) inside a $z=3$ strongly coupled plasma in $ 1+1 $ dimensions clearly exceeds the conjectured upper bound. This observation is not surprising in the context of non relativistic hydrodynamics where the Lorentz boost symmetry is explicitly broken near the UV fixed point of the theory. We further use this value of $ v_s$ obtained in (\ref{E8}) to show that the $\zeta_{lif}/s >1/4 \pi$  which is above the conjectured lower bound \cite{David}.  But we should also note that, since $\Gamma_{s}$ and $v_s$ are not independent variables as shown in (\ref{vsref}), indicating that the holographic setup considered here may not be the most general, therefore in the present calculation the value of  $\zeta_{lif}/s$ is constrained by the corresponding value of $v_s.$ %We have verified this fact independently both from the field theory (hydrodynamic) as well as the holographic calculations. Need to be changed} 
The holographic calculations provide us some exact value for this ratio which for our present gravity set up turns out to be approximately 0.1989.\par

%There are in fact some systems where the dynamical exponent ($ z $) is very close to 3 \cite{conmat}, and the specific heat measurements are also performed for them. So we  hope that the information regarding the speed of sound ($ v_s $) could in principle be estimated for fluids in $ 1+1 $ dimensions. and this opens up the possibility to test various predictions of Gauge/gravity duality in near future.
%\\
\section*{Acknowledgements} Authors acknowledge Aninda Sinha and Justin David for useful discussions.  We also thank our anonymous referees for valuable comments on the manuscript.  DR was supported through the Newton-Bhahba Fund and he would like to acknowledge the Royal Society UK and the Science and Engineering Research Board India (SERB) for financial assistance. AB acknowledges support form JSPS fellowship (P17023) and Yukawa Institute of Theoretical Physics.
%%%%%%%%%%%%%%%%%%%%
%%%%%%%%%%%%%%%%%%%%
%%%%%%%%%%%%%%%%%%%% 
%\newpage

\end{document}